\documentclass{article}

\def\be{\begin{equation}}
\def\ee{\end{equation}}
\def\bea{\begin{eqnarray}}
\def\eea{\end{eqnarray}}

\usepackage{graphics}
\usepackage{graphicx}

\begin{document}

\title{Dynamics of magnetic flux lines in the \\ 
       presence of correlated disorder}

\author{Thomas J. Bullard \\ 
        Department of Physics and Center for \\
	Stochastic Processes in Science and Engineering, \\
	Virginia Polytechnic Institute and State University, \\
        Blacksburg, VA 24061-0435, USA
\and
        Jayajit Das \\
        Department of Chemical Engineering \\ 
        and Materials Science Division, \\
        Lawrence Berkeley National Laboratory, \\
        University of California, Berkeley, CA 94720, USA
\and
        Uwe C. T\"auber \\
	Department of Physics and Center for \\
	Stochastic Processes in Science and Engineering, \\
	Virginia Polytechnic Institute and State University, \\
	Blacksburg, VA 24061-0435, USA}

\date{\today}

\maketitle

\begin{abstract}
We investigate the dynamics of interacting magnetic flux lines driven by an 
external current in the presence of linear pinning centers, arranged either in 
a periodic square lattice or placed randomly in space, by means of 
three-dimensional Monte Carlo simulations. 
Compared to the non-interacting case, the repulsive forces between the vortices
reduce the critical current $J_c$, as determined from the depinning threshold 
in the current-voltage (I-V) characteristics.
Near the depinning current $J_c$, the voltage power spectrum $S(\omega)$ 
reveals broad-band noise, characterized by a $1/\omega^{\alpha}$ power law 
decay with $\alpha \leq 2$. 
At larger currents the flux lines move with an average velocity $v_{cm}$.
For a periodic arrangement of columnar pins with a lattice constant $d$ and 
just above $J_c$, distinct peaks appear in the voltage noise spectrum 
$S(\omega)$ at $\omega \sim v_{cm}/d$ which we interpret as the signature of 
stick-slip flux line motion.
\end{abstract}

\newpage

\section{Introduction}

Magnetic vortices in the mixed state of high-$T_c$ superconductors in the
presence of defects --- either localized point pinning centers, such as oxygen 
vacancies or extended columnar pins introduced into the material via 
high-energy ion irradiation --- offer a complex system which 
has attracted the attention of both experimentalist and theorists from various
disciplines of physics in recent years \cite{BLAT}.
This area of research is also inspired by important technological questions 
triggered by the search for effective pinning centers in order to minimize 
dissipation when current is passed through the superconductor \cite{BLAT}. 
From a more fundamental viewpoint, one is interested in the non-equilibrium 
dynamics of a disordered flux line system.
When subject to an external electric current, non-trivial steady states are
generated which can be directly probed experimentally.
For example, the current-voltage (I-V) characteristics \cite{BLAT}, the voltage
noise power spectrum \cite{MAE1,MAE2,SHOBHO1}, or the local flux density noise 
\cite{MAE2} measured in experiments bear important signatures of the different 
non-equilibrium steady states. 
At low currents, transport can take place due to thermally activated 
`tunneling' of flux lines through the defects \cite{BLAT}, while at larger 
currents various intriguing steady states have been proposed, such as plastic 
flow where flux lines move in channels \cite{OLSON}. 
With increasing current the flux lines tend to reorder and move in a lattice 
(moving Bragg glass for point vortices \cite{NELSON,NATTER}), or the channels 
in the plastic flow regime can organize in a periodic structure (smectic phase)
in the transverse direction of the external current \cite{NATTER}. 

The presence of such steady states is reflected in the voltage noise spectrum;
e.g., reordering would give rise to characteristic `washboard' noise features. 
Recent experiments have in fact demonstrated the presence of both narrow- and 
broad-band noise in the measured voltage and local flux density spectra 
\cite{MAE1,MAE2,SHOBHO1}. 
Our goal is in addition to study how these noise spectra depend on the type of 
defects (correlated or point-like) present in the sample. 
However, since analytical studies of the vortex dynamics are usually limited to
asymptotic regimes \cite{BLAT,NEL,RISTE1} where the current $J$ is either much 
smaller or much larger than the critical current $J_c$, one needs to resort to 
either Langevin molecular dynamics simulations \cite{OLSON,MARC,TANG,REIC} or 
Monte Carlo techniques \cite{RYU1,RYU2} to study the various non-equibrium 
steady states which occur in the intermediate current regime.
Owing to their complexity, these numerical simulations are largely restricted 
to two dimensions \cite{OLSON,MARC,TANG,REIC} and very few three-dimensional 
studies have been reported \cite{RYU1,RYU2,BLAT2,PHYSICA,CHEN}. 
In this article we present the results of fully three-dimensional Monte Carlo 
simulations of interacting magnetic vortex lines in the presence of columnar 
defects. 
We investigate the effects of the repulsive forces between the flux lines and 
their interactions with extended defects in the I-V characteristics and the 
voltage noise spectrum, focusing on those aspects which are inaccessible by 
two-dimensional simulations. 

This article is organized as follows. 
The model is introduced in Sec. II and the Monte Carlo scheme is described in 
Sec. III. 
In Sec IV we present the results of our simulations for the I-V characteristics
and voltage noise power spectra. 
Some conclusions are offered in Sec. V.

\section{The model}

We consider a system consisting of $n$ flux lines in the London approximation 
in the presence of $N_p$ defects.
Our description utilizes an effective Hamiltonian \cite{NELSON}
\begin{eqnarray}
& & H[{\bf r}_i] = \sum_{i=1}^{n} \frac{\varepsilon}{2} \int_{0}^{L} ds
\left[ \frac{1}{\Gamma^2} \left( \frac{d{\bf r}_{i\perp}(s}{ds} \right)^2 +
\left( \frac{dz_i(s)}{ds} \right)^2 \right] \label{ham} \\
& & +\sum_{i,j=1 ; i\ne j}^{n} \epsilon_{int}\int_{0}^{L} ds \,
K_0\!\bigg(\frac{|{\bf r}_{i\perp}(s)-{\bf r}_{j\perp}(s)|}{\lambda_{ab}}\bigg)
+ \sum_{i=1}^{n} \int_{0}^{L} ds \, V_p[{\bf r}_i(s)] \ , \nonumber
\end{eqnarray}
where ${\bf r}_i(s) \equiv [{\bf r}_{i\perp}(s),z_i(s)]$ describes the 
configuration of the $i$th flux line in three dimensions, and the crystalline 
axis ${\bf c}$ of the superconducting material (as well as the external 
magnetic field) is oriented parallel to $\hat{\bf z}$, in a sample of thickness
$L$. 
The first term gives the line tension energy of a magnetic flux line.
This linear elastic form of energy holds good as long as 
$|d{\bf r}_{\perp}(z)/dz| < 1/\Gamma$ \cite{BRAN}, where $\Gamma$ denotes the 
anisotropy ratio \cite{BLAT,NEL}. 
The line stiffness is given by 
$\varepsilon\approx \varepsilon_0 \ln(\lambda_{ab}/\xi_{ab})$ \cite{BLAT,NEL}, 
with $\varepsilon_0=(\phi_0/4\pi \lambda_{ab})^2$ 
($\phi_0 = h c / 2 e$ is the magnetic flux quantum). 
$\lambda_{ab}$ and $\xi_{ab}$ denote the penetration depth and the 
superconducting coherence length in the $ab$ plane, respectively. 
For high-$T_c$ materials, $\Gamma \gg 1$ \cite{BLAT}, giving the quadratic 
form of the line tension energy a wider range of validity. 
The repulsive interaction energy between the flux lines is approximated to be
local in the $z$ coordinate, in accord with the London limit.
It is represented by the second term in (\ref{ham}), where 
$\epsilon_{int}=\phi_0^2/(8\pi^2\lambda_{ab}^2)$ and $K_0(x)$ is the modified 
Bessel function which varies as $-\ln x$ when $x \ll 1$ and 
$\sim x^{-1/2} e^{-x}$ as $x \gg 1$. 
We model the pinning potential as a sum of $N_p$ independent potential wells,
$V_p[{\bf r}(s)]=\sum_{k=1}^{N_p} U \Theta(r_p-|{\bf r}(s)-{\bf r}^{(p)}_k|)$,
where $r_p$ is the radius, $\Theta(x)$ denotes the Heaviside step function, 
and the ${\bf r}^{(p)}_k$ indicate the $N_p$ locations of the pinning sites. 
The free energy $F_L(T)$ per unit length is then defined through 
$\exp[-\beta F_L(T) L]= \int \prod_{i=1}^n D[{\bf r}_i] 
\exp \left( -\beta H[{\bf r}_i] \right)$, where $\beta=(k_B T)^{-1}$.
We study the dynamics of the flux lines when an external current ${\bf J}$ is 
applied through the system which produces a Lorentz force 
${\bf f}_L=(1/c)\hat{z} \times {\bf J}=(J/c)\,\hat{\bf x}$, per unit length of 
each flux line. 
We resort to a three-dimensional Monte Carlo (MC) simulation to deal with the 
system in an intermediate range of driving currents where the system appears to
be intractable by means of analytical methods.

\section{Monte Carlo simulation}

In our Monte Carlo simulation each flux line ${\bf r}(s)$ is modeled by 
$N=L/a_0$ points where the $i$th point is located at 
${\bf r}(i) \equiv [x(i),y(i),z(i)]$ and interacts with its nearest neighbors 
via a simple harmonic potential 
$\frac{\varepsilon}{2}\sum_{\langle j \rangle = i-1,\, i+1}\,[\Gamma^{-2} 
({\bf r}_{\perp}(i)-{\bf r}_{\perp}(\langle j\rangle))^2+(|z(i)-z(\langle j 
\rangle)|-a_0)^2]$. 
The line is placed in a box of size $L_x \times L_y \times L_z$ with periodic 
boundary conditions in all directions. The interaction energy between a pair
$(l,m)$ of flux lines is given by $\epsilon_{int} \sum_{i=1}^{N} 
K_0(|{\bf r}_{\perp l}(i)-{\bf r}_{\perp m}(i)|/\lambda_{ab})$.
To calculate the interaction energy we resort to a well-known method 
\cite{DAAN} of truncating the potential beyond a cut-off length 
$l_{cutoff} \gg \lambda_{ab}$ in order to keep the computation
time within a sensible limit. We
make sure that our results do not depend on the size of $l_{cutoff}$.  
Since we are interested in studying the effects of defect correlations in 
the dynamics, we restrict our simulation to low temperatures $T/T^{*}<1$. 
Here $T^{*}$ denotes the temperature where entropic corrections due to thermal 
fluctuations become relevant for pinned flux lines \cite{NELSON}.
The columnar pins and point defects are respectively modeled by uniform
cylindrical or spherical potential wells of (uniform) strength $U$ and radius 
$b_0$. The cylindrical wells are oriented parallel to the {\bf c} axis. 
We investigate the dynamics for three different distributions of defects, 
namely, (i) columnar pins either distributed randomly, or (ii) arranged in a 
square lattice in the $xy$ plane, and (iii) point defects distributed randomly 
in the sample. 
Since in the presence of weak currents ${\bf J}=-J\hat{\bf y}$ the flux lines 
locally move according to equilibrium dynamics, one may incorporate the effect 
of the force in the MC by introducing an additional work term 
$-\sum_{i=1}^{n}{\bf f}_L \cdot \int_{0}^{L}ds\, {\bf r}_i(s)$ in the 
Hamiltonian (\ref{ham}) \cite{NEL,RYU1,RYU2}.  

To initiate the simulation, flux lines of straight configurations parallel to 
the ${\hat {\bf z}}$ axis are nucleated at random positions in the $xy$ plane. 
We then let the system equilibrate. 
Once equilibrium has been reached we turn on the external current and let the 
system evolve.
We have checked that the results in the steady state are independent of the 
initial configurations. 
At each trial a randomly chosen point on the line is updated according to a 
Metropolis algorithm \cite{BARK}. 
The point can then move in a random direction a maximum distance of 
$\triangle < b_0/\sqrt{3}$ to guarantee interaction with every defect. 
In the simulation the values $z_i$ are held fixed; we have checked that our
results do not change qualitatively even if these positions $z_i$ are allowed 
to fluctuate \cite{PHYSICA}. 

The drift velocity of the flux lines is proportional to the average velocity of
the center of mass (CM) ${\bf v}_{cm} = n^{-1} \sum_{i=1}^{n} \langle
\overline{[{\bf R}_{i\,cm}(\tau)-{\bf R}_{i\,cm}(0)]/\tau} \rangle$, where 
${\bf R}_{i\,cm}(\tau)-{\bf R}_{i\,cm}(0)$ is the distance traversed by the CM 
of the $i$th flux line in a time interval of length $\tau$; 
$\langle\cdot\cdot\cdot\rangle$ and the overbar denote the averages over Monte 
Carlo steps (MCS) in the steady state and over different disorder realizations,
respectively. 
The voltage drop measured in experiments is caused by the induced electric 
field ${\bf E}={\bf B}\times {\bf v}_{cm}/c$ \cite{TINK}.
All length and energy scales are measured in units of $b_0$ and 
$\varepsilon_0$, respectively.
The average distance between the defects for a uniform random distribution and 
the lattice constant for the periodic array of columnar pins was taken as 
$d = 15 \, b_0$.
The parameters $\lambda_{ab}$, $\xi_{ab}$, $\varepsilon$, $U$, and $\Gamma$ 
were chosen to be $56 \, b_0$, $0.64 \, b_0$, $4 \, \varepsilon_0$, 
$0.0075 \, \varepsilon_0$, and $4$ respectively in the simulation; these 
numbers are consistent with the experimental data for high-$T_c$ materials 
\cite{BLAT}. 
Therefore thermally induced bending and wandering of the flux lines are largely
suppressed, and we can interpret our results in terms of low-temperature 
dynamics. 
In the simulations all sets of data were collected in the steady state, which 
was typically reached after $t>10^5$ MCS and $t>10^6$ MCS for columnar defects 
and point pins, respectively. 
The size of the system ranged from $60\times 9 \times 60$ to 
$2000\times 9 \times 60$ in the simulations, and the data were averaged over 
$20$ to $50$ realizations of disorder. 
The value of $\tau$ ranged from $30$ to $500$ MCS in the simulations.

\section{Results}

\begin{figure}
\includegraphics[scale=0.5]{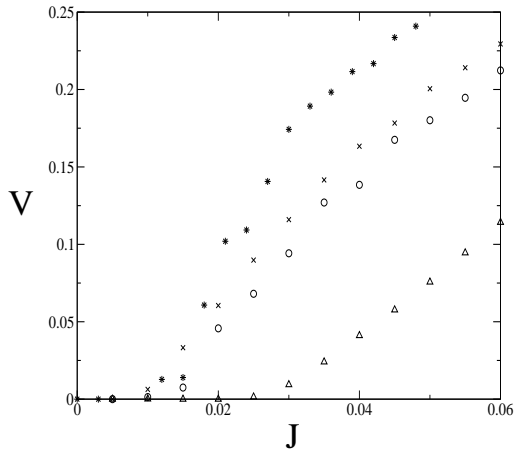}
\caption{\label{iv}
  I-V characteristics for interacting flux lines in the presence of columnar 
  defects with random ($\times$) and periodic ($\circ$) spatial distribution. 
  Note that $J_c$ is almost the same in both situations. 
  The corresponding I-V plot for a single flux line with periodically arranged
  columnar defects ($\triangle$) shows a higher $J_c$ (see text). 
  In the non-interacting (single line) case with columnar pins $J_c$ is almost 
  two times larger as compared to the interacting system. 
  The I-V plot for a single flux line with random point defect distribution 
  ($\ast$) is also given for comparison (see text).
  For currents $J>J_c$ the system with randomly distributed columnar pins 
  displays a larger resistance (higher average vortex velocity) than the one 
  with a periodic defect distribution.}
\end{figure}

In the absence of disorder and any external drive, at low temperatures 
(in our simulations $T < T^* \ll T_c$) the flux lines arrange themselves in a 
triangular Abrikosov lattice \cite{TINK}. 
In the presence of disorder the lattice structure is replaced by glass phases 
\cite{NELSON}, namely the so-called vortex glass in the case of uncorrelated 
point disorder, and the Bose glass in the presence of columnar defects
\cite{BLAT}. 
As an external current is applied the nearly localized vortices (since $T<T^*$ 
the flux lines rarely tunnel through the defects) in the glassy phases start 
moving once a critical driving current is reached.
At $T=0$, and in the absence of mutual vortex interactions, the flux lines are 
strictly bound to one or more pins in equilibrium.
At the critical depinning current $J=J_c$ one then encounters a continuous
non-equilibrium phase transition \cite{ONU,KAR1,KAR2}.
This zero-temperature phase transition becomes rounded to a crossover at
$T > 0$, but a characteristic `critical' current separating the localized and
moving vortex regimes can still be inferred at sufficiently low temperatures,
as seen in the I-V characteristics displayed in Fig.~\ref{iv}, which were 
computed at $T=0.25\times 10^{-3} \varepsilon \ll T^{*}$.

It turns out that the critical current $J_c$ is actually higher for a system
with columnar pins as compared to point disorder because of the increased 
defect correlations \cite{PHYSICA}.
However the value of $J_c$ is the same for both random and periodic spatial 
distribution of the pins \cite{PHYSICA}. 
Repulsive interactions between the flux lines facilitate flux creep which is 
reflected in the delocalization of the vortices at a considerably lower current
than for the non-interacting case (Fig.~\ref{iv}). 
The I-V characteristics also reveal that the voltage at the same current is 
higher for the system with a random as compared to periodic defect 
distribution. 
Similar behavior is observed in the dynamics of the single flux line, and can 
be qualitatively explained by an effectively faster transit time in the random 
pin distribution compared to the periodic columnar defect array \cite{PHYSICA}.
We have also found that in the system with a single flux line in the presence 
of a periodic pin arrangement the I-V characteristics depends strongly on the 
orientation of {\bf J} with respect to the lattice direction, for this 
determines the effective density of defects encountered by the moving vortex. 
A similar feature is likely to survive in the presence of interactions.

\begin{figure}
\includegraphics[scale=0.5]{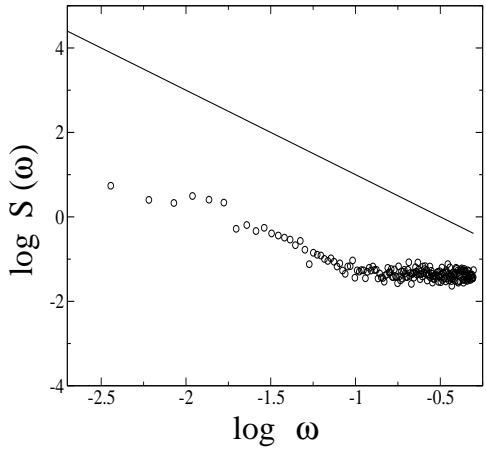}
\caption{\label{cnoi}
  Log-log plot of the velocity noise power spectrum $S(\omega)$ vs. frequency 
  $\omega$ for a system with columnar defects near the depinning threshold at 
  $J\approx J_c$. 
  The solid line, given as a guide, has slope $-2$.}
\end{figure}

The velocity (proportional to the induced voltage) noise is calculated by 
computing the velocity fluctuations about ${\bf v}_{cm}$. 
We evaluate the power spectrum 
$S(\omega)=\overline{\tilde{v}_x(\omega)\tilde{v}^{\ast}_x(\omega)}$, where 
$\tilde{v}_x(\omega)$ is the Fourier transform of the velocity fluctuation 
$\tilde{v}_x= n^{-1} \sum_{i=1}^{n} [v_{ix}-\langle v_{ix} \rangle]$, with 
$v_{ix}(t)=[X_{i\,cm}(t+\tau)-X_{i\,cm}(\tau)]/\tau$ being the center-of-mass
velocity of the $i$th flux line. 

Near $J_c$ we observe a power-law decay in the voltage noise spectrum:
$S(\omega) \sim 1/\omega^{\alpha}$, where an effective exponent 
$\alpha\approx 2$ is observed for almost a frequency decade in the system with 
periodically arranged columnar pins (Fig.~\ref{cnoi}).
At low frequencies there are indications that the effective exponent may be
smaller than $2$.
For a single line in the presence of random point defects and at $T=0$, the 
value of $\alpha$ can be determined through a functional renormalization group 
calculation which gives $\alpha=1.5$ to one-loop order in three dimensions 
\cite{KAR1,KAR2}, with corresponding mean-field value $\alpha=2$.
The power law behavior observed in the simulation (Fig.~\ref{cnoi}) can be 
interpreted as a remnant of the zero-temperature depinning transition.
In experiments \cite{SHOBHO1}, such broad-band noise (BBN) features show a 
$\omega^{-\alpha}$ decay with $\alpha\approx 2$.

\begin{figure}
\includegraphics[scale=0.88]{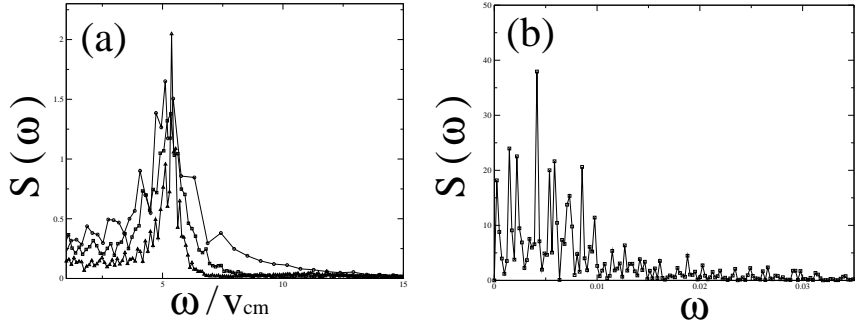}
\caption{\label{pernoi} 
  (a) Plot of the velocity noise power spectrum $S(\omega)$ vs. $\omega/v_{cm}$
  for a system with columnar pins arranged in a periodic array for $J=0.05$ 
  (circles), $J=0.07$ (squares), and $J=0.09$ (triangles). 
  The peaks occur at the same scaled frequency 
  $\tilde{\omega}\propto v_{cm}/d$.
  Note that the width of the peak decreases as $J$ is increased indicating the 
  wider separation between the time scales $\tau_r$ and $\tau_v$.
  (b) Plot of $S(\omega)$ vs. $\omega$ for columnar pins distributed randomly 
  in the sample for $J=0.05$ (squares). 
  Notice the presence of secondary peaks in the spectrum (see text).}
\end{figure}

For currents $J>J_c$, three distinct time scales can appear in the dynamics:
the residence time $\tau_r$ which is the duration a vortex line is trapped on 
the defects, $\tau_v$, the time taken by the flux line to travel between
the pins, and $\tau_w$, the time taken by the moving vortex lattice in the 
event of reordering to traverse a distance of the lattice constant.
In the regime where these time scales are well separated and $\tau_r<\tau_v$, 
as well as $\tau_r<\tau_w$, we can expect $\tau_v$ and $\tau_w$ to produce 
pronounced peaks in the velocity noise spectrum. 
We first summarize the results from the non-interacting case where the time 
scale $\tau_w$ is entirely absent (stick-slip motion). 
In Fig.~\ref{pernoi}(a) we plot $S(\omega)$ at different driving currents for a
periodic distribution of columnar pins.
The spectra reveal a single peak corresponding to the frequency 
$\tilde{\omega} \propto v_{cm}/d \propto 1/\tau_v$. 
Note that the width of the peak decreases as the external current ${\bf J}$ 
increases, which widens the time scale separation between $\tau_r$ and 
$\tau_v$. 
In a random defect distribution, see Fig. \ref{pernoi}(b), there would be a 
distribution of the time scales $\tau_v$ around the average value $d/v_{cm}$. 
Therefore there could be many secondary peaks present in the spectrum arising 
due to the distribution of $\tau_v$ in addition to the principle peak which 
corresponds to $\tau_r\propto d/v_{cm}$. 
These secondary peaks would eventually diminish in intensity as the 
distribution of the time scales will narrow down at $d/v_{cm}$ with the results
averaged over more defect configurations. 
Unlike for correlated defects, the time scales $\tau_r$ and $\tau_v$ are not 
well separated for point pins and the peaks turns out to be considerably 
suppressed as compared to the situation for columnar defects \cite{PHYSICA}.

\begin{figure}
\includegraphics[scale=0.5]{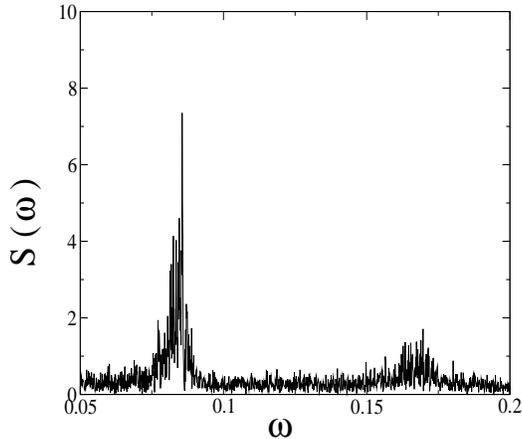}
\caption{\label{ints}
  Plot of the velocity noise spectrum $S(\omega)$ vs. $\omega/v_{cm}$ for a
  system of periodically arranged columnar pins at $J=0.07$ with interacting 
  flux lines. 
  The primary peak corresponds to the rescaled frequency 
  $\tilde{\omega}\propto v_{cm}/d$.}
\end{figure}

The presence of interactions between the flux lines may introduce the third 
time scale $\tau_w$ if the vortex lines align themselves into a lattice 
structure at large external currents. 
For example the predicted moving vortex lattice (in the presence of point
defects) \cite{NATTER} would give rise to a distinct 
$\tau_w\propto a_m/v_{cm}$, where $a_m$ is the lattice constant of the moving 
flux line array. 
We have not yet detected any signature of the corresponding narrow-band noise 
in the presence of columnar defects with periodic spatial distribution, even 
when the average distance between the flux lines is much less than the 
periodicity of the pins. 
This implies absence of reordering in the observed range of external currents. 
We do find a peak in $S(\omega)$ corresponding to $\tau_v\propto d/v_{cm}$ 
(Fig.~\ref{ints}) which is also observed in the non-interacting case. 
However we might expect reordering when the density of the flux lines is larger
than the defect density because then the effect of defect correlations becomes
masked.
 
We have noticed a marked difference in $S(\omega)$ between columnar and point 
pins which results from the lack of spatial correlation in the $z$ direction 
for the point disorder.
Notice that this distinction can be observed only in a three-dimensional 
dynamical simulation. 
It is due to a generic difference between correlated and localized pinning
potentials.
The ensuing absence or presence of the narrow-band noise peaks in the voltage 
noise spectrum $S(\omega)$, and its specific features, could perhaps be 
utilized as a signature to identify and characterize the type of disorder 
present and responsible for flux pinning in experimental superconducting 
samples as well. 

\section{Conclusion} 

In conclusion we have investigated the three-dimensional dynamics of 
interacting flux lines in the presence of correlated defects, either arranged 
in a periodic array or placed randomly in space. 
We find that the nature of the current-voltage characteristics depends on the 
defect correlation as well as the spatial distribution of the pinning centers. 
The voltage noise spectrum shows pronounced peaks in the case of correlated 
pins. 
The difference in the power spectra at $J>J_c$ between columnar and point 
disorder may serve as a novel diagnostic tool to characterize the pinning
centers in superconducting samples.

\section{Acknowledgements}
We would like to thank S. Bhattacharya, A. Maeda, B. Schmittmann, and
R. K. P. Zia for valuable discussions. 
This research has been supported by the National Science Foundation
(grant no. DMR 0075725) and the Jeffress Memorial Trust (grant no. J-594).
JD is supported by a Department of Energy grant through Lawrence Berkeley 
National Laboratory.

\end{document}